\documentclass[a4paper,12pt,english,german]{article}

\usepackage{mathtools}
\usepackage{graphicx}
\usepackage{amssymb,amsmath,scalefnt,cite}
\usepackage{titlesec}

\usepackage{hyperref}
\usepackage{etoolbox}
\appto\appendix{\counterwithin{equation}{section}}

\begin{document}

\centerline{\bf Entropy dynamics for a harmonic propeller-shaped\\}
\centerline{\bf  planar molecular quantum Brownian rotator}

\bigskip

\centerline{J. Jekni\' c-Dugi\' c$^1$, I. Petrovi\' c$^2$, M. Arsenijevi\' c$^3$\footnote{Author to whom any correspondence should be addressed at the e-mail momir.arsenijevic@pmf.kg.ac.rs}, M. Dugi\' c$^3$}

\bigskip

$^1$University of Ni\v s, Faculty of Science and Mathematics, \\ \hangindent2em Vi\v
segradska 33,  18000 Ni\v s, Serbia

$^2$Svetog Save 98, 18230 Sokobanja, Serbia

$^3$University of Kragujevac, Faculty of Science, \\ \hangindent2em Radoja
Domanovi\' ca 12, 34000 Kragujevac, Serbia

\begin{abstract}
Dynamic stability is vital for the operation and control of molecular
nanodevices, particularly molecular cogwheels in an external harmonic
field. We used the relative change in rotator entropy as a measure of
dynamical stability. This rotator dynamic is described by the Caldeira-Leggett
quantum master equation: the  main challenge is the absence
of a general solution for this equation. As an alternative, we look for
the initial states (pure or mixed) of the rotator that result in
low entropy change. In order to create candidate initial states, we use an ansatz
that align with the maximum entropy principle. We then find conditions
for state preparation and system/bath parameters to achieve
a small relative change in both linear and differential entropy in underdamped
and non-underdamped regimes. We identify cases with
relatively small entropy change and discuss their practical feasibility.
\end{abstract}

\section{\label{sec:SectionI}Introduction}

Propeller-shaped molecular cogwheels are among the basic elements of  molecular nano-machinery \cite{Jones,Godsel,Kotas}. Their complex behavior requires a multidisciplinary approach \cite{Soe1,pregled,KNJIGA}. In this regard, theoretical studies are often instrumental. Classical-physics models are typically effective, particularly for larger molecular species at higher temperatures. However, there are instances where classical models may be insufficient \cite{Kotas}. Specifically, external operations on molecules are frequently performed in short time intervals. In these cases, quantum-mechanical corrections can build up and become important before the system relaxes or thermalizes. Recently, systematic research has begun \cite{JJD1,JJD2}. These  include studies on decoherence \cite{Stickler,Zhong,Stikler}, diffusion \cite{Stickler,Stikler,Fjuri}, friction \cite{Stickler,Stikler,Ju1}, and relaxation or thermalization \cite{Stikler,Srinski,Kamba}. The potential applications of these researches  are broad. They include cooling nanoscale rotators \cite{Zhong,cooling1,cooling2}, orientational quantum metrology \cite{metrology1,metrology2}, torque sensors \cite{Ju1,sensors,Ahn}, nano-g acceleration sensitivity \cite{acceleration}, quantum heat engines \cite{heatengine}, qubit construction \cite{QUBIT}, and foundational topics like experimental tests of objective collapse models \cite{bera,collapse} or testing quantumness in massive objects \cite{das}.

Dynamic stability is a necessary condition for controlling and operating     molecular nano-setups. In our recent work, we used the standard deviation and the mean first passage time to evaluate the rotational stability of molecular cogwheels (MC) \cite{JJD1,JJD2}. These MCs were modeled as Caldeira-Leggett-type planar quantum Brownian rotators. The number of blades, $n$, is key to  defining the MC system and adds a new free parameter to the standard Caldeira-Leggett (CL) model \cite{JJD1,JJD2}. This extension preserves the fundamental  structure of the original CL model while introducing a single additional parameter that accounts for the system’s finite size. Thus, the parameter $n$
acts as a new degree of freedom, enabling physical behavior that does not appear in the analyses based on the traditional model. The results show that optimization of the rotator's parameters and procedures is needed for a reliable function of the rotator. There are, of course, many other criteria that could be useful and therefore warrant further investigation. The practical reliability of these criteria can only be determined in experiments.

In this study,  entropy was  used  as the  standard measure of dynamic stability. A smaller entropy change indicates a higher stability of  molecular cogwheel rotations \cite{ZUREK,Brojer}. The literature shows many results on entropy dynamics \cite{ZUREK,BENATI,SANTOS}, quantum correlation dynamics \cite{EISERT}, and the purity or stability of open-system states \cite{HORNBERGER,DATA,GU}. It also covers quantum Markovianity \cite{ALIPUR,BROJER}. A number of the related approaches investigate the non-Hermitian models, their quantum simulations and the information change as well as dynamical change of the generalized entropies, such as the Renyi entropy
\cite{Rost,Zheng,Sergi,LiZheng,LiuZheng}.

Ideally,    zero entropy change occurs when the rotator starts in a stationary state or moves between pure states. The decoherence-induced pointer basis offers stable or "robust" states for the open system. As shown below, our considerations go beyond these well-known results.

We study the main forms of the Caldeira-Leggett model \cite{Brojer,CL}--the standard non-Markovian model and its two Markovian limits (the Markovian limit and the decoherence limit \cite{Brojer}). To align with past studies \cite{JJD1,JJD2}, we do not limit the system's parameters, except for a moderate temperature $T$ \cite{GARDINER} and the weak coupling ($\gamma_{\circ}\ll 1$) with the environment, assuming a small rotation angle. We omit very small rotations from our analysis. While this would simplify the model to the standard Cartesian form, such small rotations are often hard to observe and control. Thus, we use the angle-angular momentum formalism. We find that a transition from one pure state to another (the pure pure-state dynamics or PPSD condition \cite{Dugic}) does not occur in the processes we analyze. Using the Maximum Entropy Principle (MEP) \cite{DILL}, we identify two exact stationary states for the Brownian particle in an external harmonic field. We recognize the standard mixed state \cite{Brojer} and also discover another exact stationary state (often overlooked), which is pure. Based on this, we suggest an {\it alternative approach} to the problem.

We construct the initial rotator states that meet the MEP. This principle expects these states to show a small entropy change. To address the CL processes, we suggest the following approach: we use the known, time-dependent standard deviations for the angle and angular momentum operators \cite{JJD1} for the stationary states. Next, we calculate both differential and linear entropy (which can be negative) \cite{Gibson}. We look for parameter values that yield a small entropy change ($\vert\delta S\vert\equiv\vert S(t)-S(0)\vert/S(0)$) in both the underdamped and non-underdamped regimes. The rotator's self-energy, $E$, defines the regimes, characterized by the inequalities $n_{max}\hbar\gamma_{\circ} \ll E < k_BT$ and $n_{max}\hbar\gamma_{\circ}\ll k_BT < E$, respectively. Here, $n_{max}$ is the maximum number of blades of the rotator and $E\sim\hbar\omega$.

The structure of this paper is as follows. Section~\ref{sec:SectionII} demonstrates the impossibility of PPSD and the unavoidable increase of entropy for all considered processes. Section~\ref{sec:SectionIII} proposes the MEP as a guide for finding the correct initial states of the harmonic Brownian rotator. Based on this, we construct a pair of states--one pure and one mixed--and calculate the temporal entropy change for both linear and differential entropy across all processes of interest in the two regimes. Results are partially obtained numerically and presented in Section~\ref{sec:SectionIV}. Section~\ref{sec:SectionV} offers a discussion, and we conclude in Section~\ref{sec:SectionVI}.

\section{\label{sec:SectionII}The pure state dynamics}

The transition from one pure state to another creates a trajectory $\vert\Psi(t)\rangle$ in the system's Hilbert state space. This transition keeps entropy at zero at all times. When dynamics are linear and deterministic, they become predictable and stable. However, there is no general rule in this regard, even  for the  Markovian open systems \cite{Dugic}. Thus, we must check for dynamical stability in each specific model. If we can identify these states, preparing the open system in such states will help us achieve our goal. It will ensure dynamical stability through the predictability of MC dynamics.

In this paper, we focus on the standard CL process. This process is non-Markovian and even non-completely positive (non-CP) for very short time intervals. We also look at two Markovian variations of the CL process. All these processes belong to the C class of dynamical maps \cite{Invertibility}.

We consider small rotations described by the azimuthal angle $\hat\varphi$ around the $z$-axis. We also include the conjugate angular momentum $\hat L_z$ of the cogwheel. The cogwheel is subjected to a harmonic external potential with circular frequency $\omega$. This leads to the Hamiltonian $\hat H=\hat L_z^2/2nI_{\circ}+nI_{\circ}\omega^2\hat\varphi^2/2$. Here, $I_{\circ}$ and $\gamma_{\circ}$ represent the average moment of inertia and damping factor for the number $n\in[1, 10]$ of the cogwheel blades.

The standard CL process is described by the master equation for the molecular cogwheels (in the Schr\" odinger picture) \cite{JJD1,JJD2}:
\begin{equation}\label{eqn:novo1}
{d\hat\rho\over dt} = -{\imath\over\hbar} [\hat H_r,\hat\rho] - {\imath n\gamma_{\circ}\over\hbar}[\hat\varphi,\{\hat L_z,\hat\rho\}] -{2n^2I_{\circ}\gamma_{\circ}k_BT\over\hbar^2}[\hat\varphi,[\hat\varphi,\hat\rho]].
\end{equation}

\noindent while the Markovian CL processes are of the general Lindblad form
\begin{equation}\label{eqn:novo2}
{d\hat\rho\over dt} = -{\imath\over\hbar}[\hat H_r,\hat\rho] +\sum_i\Gamma_i\left(\hat A_i\hat\rho\hat A^{\dag}_i-{1\over 2}\{\hat A^{\dag}_i\hat A_i,\hat\rho\} \right),
\end{equation}

\noindent where $\hat H_r$ is the open system's renormalized Hamiltonian and $\hat A_i$ stands for the Lindblad operators that induce non-unitary dynamics with the damping factors $\Gamma_i$; $k_B$ is the Boltzmann constant and $T$ the bath temperature. The single Lindblad operators of interest for the Markovian processes read $ \sqrt{{4nI_{\circ}k_BT/\hbar^2}}\hat\varphi + \imath\sqrt{{1/ 4nI_{\circ}k_BT}}\hat L_z$ for the ''Markovian (very high temperature) limit'' and $\hat\varphi$ for the ''decoherence (recoilless) limit'' of the process Eq. (\ref{eqn:novo1}) with the damping factors $n\gamma_{\circ}$ and $-{2n^2I_{\circ}\gamma_{\circ}k_BT\over\hbar^2}$, respectively.

The condition of purity, $\hat\rho^2(t)=\hat\rho(t),\forall{t}$, implies
\begin{equation}\label{eqn:novo3}
tr\left(\hat\rho(t){d\hat\rho(t)\over dt}\right)=0,\forall{t}.
\end{equation}

\noindent Substituting $d\hat\rho/dt$ into Eq. (\ref{eqn:novo3}) provides the purity condition  (the PPSD condition \cite{Dugic}) for a process that applies to all the time-local master equations.
The condition Eq. (\ref{eqn:novo3}) takes a specific form for the Markovian processes \cite{Dugic}:
\begin{equation}\label{eqn:8}
\sum_i \langle \hat A_i^{\dag}\hat A_i\rangle - \langle\hat A_i\rangle\langle\hat A_i^{\dag}\rangle = 0.
\end{equation}

After some simple algebra and the use of the formal commutator $[\hat\varphi,\hat L_z]=\imath\hbar$ follows the PPSD condition for the standard Caldeira-Leggett process:
\begin{equation}\label{eqn:7}
(\Delta\hat\varphi)^2 = {\hbar^2\over 4nI_{\circ}k_BT}.
\end{equation}

\noindent The time-fixed value of the standard deviation in Eq. (\ref{eqn:7}) is in disagreement with the known, time-dependent standard deviation \cite{Brojer}.
In effect, the standard CL master equation (\ref{eqn:novo1}) does not allow for any pure state dynamics.

For the Markovian-limit CL process follows the PPSD condition:
\begin{equation}\label{eqn:9}
{4nI_{\circ}k_BT\over\hbar^2}(\Delta\hat\varphi)^2 + {1\over 4nI_{\circ}k_BT}(\Delta\hat L_z)^2 -1=0, \forall{t}.
\end{equation}

\noindent Let us consider the short-time limit of Eq.(\ref{eqn:9}). Bearing in mind the dimensionless expressions of the standard deviations ($\sigma_{\varphi}\equiv\Delta\hat\varphi$, $\sigma_L\equiv\Delta\hat L_z/I_{\circ}\omega$ and $\sigma\equiv (\langle\{\hat\varphi,\hat L_z\}\rangle-2\langle\hat\varphi\rangle\langle\hat L_z\rangle)/I_{\circ}\omega$,
while assuming the weak-coupling limit $\gamma_{\circ}/\omega\ll 1$) and placing the additional Markovian terms to the known expressions \cite{JJD1} yields

\begin{align}\label{eqn:10}
\begin{split}
\sigma_{\varphi}^2(t)& = e^{-2n\gamma_{\circ}t} \left(\sigma_{\varphi}^2(0)\cos^2\omega t + {\sigma(0)\over 2n}\sin2\omega t + {\sigma_L^2(0)\over n^2}\sin^2\omega t\right)+\\&
+ \left({k_BT\over n I_{\circ}\omega^2} - {\hbar^2\over 16 n I_{\circ}k_BT}\right)(1-e^{-2n\gamma_{\circ}t}),\\
\sigma_{L}^2(t)& = e^{-2n\gamma_{\circ}t} \left(\sigma_L^2(0)\cos^2\omega t - {n\over 2}\sigma(0)\sin2\omega t + n^2\sigma_{\varphi}^2(0)\sin^2\omega t\right)+\\&
+ \left( {n k_BT\over I_{\circ}\omega^2}-{n\hbar^2\over 16I_{\circ}k_BT}\right)(1-e^{-2n\gamma_{\circ}t}).
\end{split}
\end{align}

\noindent Hence the short-time ($n\gamma_{\circ} t\ll\omega t\ll 1$)
\begin{align}\label{eqn:11}
\begin{split}
\sigma_{\varphi}^2(t)&\approx (1-2n\gamma_{\circ}t)\sigma_{\varphi}^2(0)+ {2\gamma_{\circ}k_BT\over I_{\circ}\omega^2}t -{\hbar^2\gamma_{\circ}\over 8I_{\circ}k_BT}t,\\
\sigma_L^2(t)&\approx (1-2n\gamma_{\circ}t)\sigma_{L}^2(0) + {2n^2\gamma_{\circ}k_BT\over I_{\circ}\omega^2}t -{n^2\gamma_{\circ}\hbar^2\over 8I_{\circ}k_BT}t.
\end{split}
\end{align}
\noindent On the use of Eq. (\ref{eqn:11}), equation (\ref{eqn:9}) gives the condition for PPSD, $8(k_BT/\hbar\omega)^2-1/32(\hbar\omega/k_BT)^2=2$, which is in sharp distinction from the initial assumption of the very high temperature.
Therefore neither the Markovian-limit process  admits a pure state dynamics.

Finally, the PPSD condition for the decoherence-limit CL process (which is also Markovian \cite{Brojer,Invertibility,Rivas}) reads:
\begin{equation}\label{eqn:12}
\sigma_{\varphi}^2=\langle\hat\varphi^2\rangle-\langle\hat\varphi\rangle^2=0, \forall{t}.
\end{equation}

\noindent Since Eq. (\ref{eqn:12}) distinguishes the un-normalizable eigenstates of the angle observable $\hat\varphi$, there is not even a single trajectory in the (separable) Hilbert state space for the rotator.

We conclude that PPSD is not allowed for any of the considered variants of the Caldeira-Leggett master equation.
Hence, the initially pure state of the rotator becomes mixed in the very next time instant thus distinguishing a necessary rise of entropy of the rotator state $\hat\rho(t)$.

\section{\label{sec:SectionIII}Quantum states from the maximum entropy principle}\label{SectionIII}

The maximum entropy principle (MEP) is important in information theory, statistical inference, and statistical mechanics \cite{DILL}. Given certain constraints, it offers a probability density with the highest entropy. This is the opposite of what we discussed in Section~\ref{sec:SectionII}. Thus, a state following the MEP should show a small change in entropy. We can take advantage of this property.

It is well known that MEP returns the normal distribution for the preassigned standard deviation ($\sigma_x$) and the mean ($x_{\circ}$) for the variable $x\in(-\infty,\infty)$ of interest:
\begin{equation}\label{eqn:13}
\rho(x)={1\over\sigma_x\sqrt{2\pi}} e^{-{(x-x_{\circ})^2\over 2\sigma_x^2}}.
\end{equation}

\noindent Equation (\ref{eqn:13}) directly gives the probability density for the classical systems. Our task is to find or propose the proper quantum-mechanical states $\hat\rho(t)$ whose kernels $\rho(x,x',t)$ return the probability density Eq.(\ref{eqn:13}), {\it such that} the found states are allowed by the Caldeira-Leggett\ processes  or at least by one of them. Given such states we can investigate the entropy dynamics for the considered processes.

In order to keep symmetrical roles of $x$ and $x'$, we choose the following, sufficiently general \cite{Brojer,Ferialdi,Trus} form of the kernel:
\begin{equation}\label{eqn:14}
\rho(x,x',t) = \mathcal{N} e^{-\alpha(x^2+x'^{2})-\beta xx'},
\end{equation}

\noindent with the normalization factor $\mathcal{N}$ and the constraint $2\alpha+\beta=(2\sigma_x^2)^{-1}$. Certain special cases to be investigated are at hand. First, for $\alpha=(4\sigma_x^2)^{-1}$ and $\beta=0$ we have the kernel for the pure (minimal-uncertainty) state $\psi(x) = \sqrt{\mathcal{N}}\exp(-(x-x_{\circ})^2/4\sigma_x^2+\imath px/\hbar)$:

\begin{equation}\label{eqn:15}
\hat \rho = \mathcal{N}\int dxdx' \exp\left(-{(x-x_{\circ})^2+(x'-x_{\circ})^2\over 4\sigma_x^2} +{\imath p(x-x') \over\hbar} \right)\vert x\rangle\langle x'\vert.
\end{equation}

\noindent Second, the choice of $\alpha=1/8\sigma_x^2+\sigma_p^2/2\hbar^2$ and $\beta=1/4\sigma_x^2-\sigma_p^2/\hbar^2$ corresponds to the mixed stationary state of the standard Caldeira-Leggett equation (\ref{eqn:novo1}) for the harmonic oscillator \cite{Brojer}:

\begin{equation}\label{eqn:16}
\hat R= \mathcal{N}\int dxdx' \exp\left(-{1\over 2\sigma_x^2}\left({x+x'\over 2}\right)^2-{\sigma_p^2\over 2\hbar^2}(x-x')^2 \right) \vert x\rangle\langle x'\vert,
\end{equation}

\noindent where $\sigma_p^2=m k_BT$ denotes the standard deviation for the momentum observable $\hat p$, $[\hat x,\hat p]=\imath\hbar$, $\mathcal{N}={1\over\sigma_x\sqrt{2\pi}}$ and $\sigma_x^2=k_BT/m\omega^2$ for the oscillator of the mass $m$ and circular frequency $\omega$, with the mean $x_{\circ}=0$. For both states the covariance function $\sigma = \langle\hat x\hat p+\hat p \hat x \rangle - 2\langle\hat x\rangle\langle\hat p\rangle=0$.

The mixed state in Eq. (\ref{eqn:16}) is known as a stationary state for the standard CL equation (\ref{eqn:novo1}) \cite{Brojer}. At very high temperatures, this state is nearly stationary for the Markovian process, but not for the decoherence limit CL process. The stationary nature of state (\ref{eqn:16}) aligns with the unavoidable entanglement that forms between the system and environment when the initial state is a pure Gaussian \cite{EISERT}.

In Appendix \ref{AppendixA}, we show that the pure state in Eq. (\ref{eqn:15}) is another {\it exact} stationary state for {\it specific choices} of system parameters. This state is also nearly stationary for the Markovian process, but not for the decoherence-limit master equation. From Eq. (\ref{eqn:9}), if we ignore the Markovian terms, the stationary values of $\sigma^2_{\varphi}=k_BT/nI_{\circ}\omega^2$ and $\sigma^2_L=nI_{\circ}k_BT$ emerge.

Since the CL processes are Gaussian \cite{Ferialdi} and the density probability in Eq. (\ref{eqn:13}) is also Gaussian, we propose a Gaussian {\it ansatz}: we will replace the time-independent standard deviations in Eqs. (\ref{eqn:15}) and (\ref{eqn:16}) with the known time-dependent standard deviations \cite{JJD1} for each process. We then use a Gaussian form for the initial time-dependent states $\hat\rho$ and $\hat R$. According to the MEP, these states share a common time-dependent probability density that is independent of $\sigma_p$ for all CL processes. By using the known standard deviations \cite{JJD1}, we obtain "equipped" time-dependent states that match the CL processes of interest. This allows us to compare them with results for other quantifiers of molecular rotation stability \cite{JJD1}.

By swapping $x$ with $\varphi$ and $p$ with $L$, we normalize the density distribution $\rho(\varphi,t)$ over the range $\varphi\in[-\pi,\pi]$. This leads us to question if $\sigma_{\varphi}$ in Eqs. (\ref{eqn:15}) and (\ref{eqn:16}) remains the standard deviation, especially in time-dependent cases. As shown in Appendix \ref{AppendixB}, this holds true if $\sigma_{\varphi}(t)\le 1$. Certain system and state parameters can satisfy this constraint (see Section~\ref{sec:SectionV} and Appendix \ref{AppendixB}). The time-independent probability distribution in Eq. (\ref{eqn:13}) represents the formal asymptotic form of the time-dependent distribution. This is consistent with the time-dependent states $\hat\rho(t)$ and $\hat R(t)$ as per the MEP. Appendix \ref{AppendixB} shows that the time-dependent probability distribution for constructed states is similar to the time-independent distribution in Eq. (\ref{eqn:13}) for exact stationary states. This similarity holds true even over short time intervals, especially in the non-underdamped regime and with more blades. Therefore, we can expect a small entropy change for the time-dependent states $\hat\rho(t)$ and $\hat R(t)$ as the initial rotator states.

\section{\label{sec:SectionIV}Entropy dynamics for the Caldeira-Leggett processes}

We examine MEP as a general criterion, focusing on the states $\hat\rho(t)$ and $\hat R(t)$. We do not impose limits like those found in CP processes \cite{BENATI}, the "predictability sieve" \cite{ZUREK}, or the "dark states" area \cite{SANTOS}. We treat these time-dependent states $\hat\rho(t)$ and $\hat R(t)$ as the initial states of the rotator.

Next, we calculate the absolute value of relative entropy change: $\vert\delta S\vert=\vert S(t)-S(0)\vert/S(0)$. We consider both linear and differential entropy for all processes. Our calculations rely on the assumption \cite{JJD1,JJD2} of very small rotation angles. We achieve this by looking at short time intervals and using numerical integrations.

According to Eq. (\ref{eqn:13}), differential entropy (DE) gives results applicable to both pure and mixed states (\ref{eqn:15}) and (\ref{eqn:16}). Since the linear entropy for the pure initial state is zero, it is not useful here. Instead, we apply the "purity loss" (see Eq. (\ref{eqn:novo3})) as a measure for the entropy change of the initially pure time-dependent state.

We select parameter values as follows: $\omega=1\gg\gamma_{\circ}=0.001, \hbar=1$ and $n\in[1,10]$. The values of $k_BT$ and $I_{\circ}$ differ based on the CL processes and dynamic regimes (underdamped and non-underdamped). Lastly, our choice of initial states must meet the conditions $\sigma_{\varphi}(t)\le 1$ and $\Delta\hat\varphi\Delta\hat L_z\ge\hbar/2$.

\subsection{\label{sec:SectionIV2}Purity loss}

Entropy is zero for every pure state. As shown in Section~\ref{sec:SectionII}, initially pure states quickly become mixed. The change in entropy can be understood as a loss of initial purity: the greater the loss of purity, the greater the entropy change when transitioning from a pure to a mixed state. In Section~\ref{sec:SectionII} we implicitly introduced a quantitative criterion for the purity loss, $\dot{p}=2tr(\hat\rho d\hat\rho/dt)$, which for the initially pure state Eq.(\ref{eqn:15}) reads:
\begin{align}\label{eqn:17}
\dot{p}_{stand}(t)&=2\left( n\gamma_{\circ} - {4n^2\gamma_{\circ}I_{\circ}k_BT\over\hbar^2}(\Delta\hat\varphi(t))^2_{stand}\right)\\ \nonumber
\dot{p}_{markov}(t)&=-2\left({4nI_{\circ}k_BT\over\hbar^2}(\Delta\hat\varphi(t))^2_{markov} + {1\over 4nI_{\circ}k_B T}(\Delta\hat L_z(t))^2_{markov}-1 \right)\\ \nonumber
\dot{p}_{decoh}(t)&=-{8n^2I_{\circ}\gamma_{\circ}k_BT\over\hbar^2}(\Delta\hat\varphi(t))^2_{decoh},
\end{align}

\noindent for the standard, Markovian and the decoherence-limit CL processes, respectively; in Eq. (\ref{eqn:17}), $(\Delta\hat\varphi)^2=\sigma_{\varphi}^2$, and $(\Delta\hat L_z)^2=(I_{\circ}\omega)^2\sigma_L^2$.
The expected negativity of the purity loss is clear in the decoherence process, which is unital \cite{Lidar}. For the standard process, the condition $\sigma_{\varphi}^2\ge \hbar^2/4nI_{\circ}k_BT$, known as the positivity condition \cite{GU,DiosiCL} implies negativity. Similarly, the negativity of the purity loss can be easily demonstrated for the Markovian CL process, consistent with the "predictability sieve" criterion \cite{ZUREK}.

For the standard process in Eq. (\ref{eqn:novo1}), we know from Appendix \ref{AppendixA} that the pure state in Eq. (\ref{eqn:15}) is a stationary state for certain parameter choices. With temperature $T$, the choice of $nI_{\circ}=\hbar^2/4k_BT$, $\omega=2k_BT/\hbar$ and $L=\langle\hat L_z\rangle=0$ leads to a standard deviation of $\Delta\varphi\equiv\sigma_{\varphi}=1$. This results in $\dot{p}_{stand}=0$, creating a recipe for zero purity loss: according to Eq. (\ref{eqn:7}), initial state prepared with  $\sigma_{\varphi}=1$ remains dynamically intact.

There is some flexibility in choosing parameter values. For propellers with average moment of inertia $I_{\circ}$, we can adjust the number of blades and the external harmonic field frequency $\omega$. In practice, we can only approximate state preservation, but we can still achieve a low entropy change.

In the Markovian-limit process, high temperatures usually lead to significant purity loss. However, selecting a small $\sigma_{\varphi}$ for the initial state in Eq. (\ref{eqn:15}) may help reduce this loss. For the decoherence CL process, a sufficiently small $\sigma_{\varphi}$ results in a relatively low entropy change.

\subsection{\label{sec:SectionIV3}Differential entropy}

For the linear variable $x\in(-\infty,\infty)$, the differential entropy simplifies to the standard deviation $\sigma_x$ \cite{Garbac}. However, we cannot apply this result directly to our situation. We set the initial expectation value $\langle\hat\varphi\rangle=0$ and choose a geometry where the total interval of $2\pi$ corresponds to the integration interval $[-\pi,\pi]$. We assume small rotations, which occur over very short time intervals. We also take very small initial values for $\sigma_{\varphi} $, similar to classical values, resulting in negative differential entropy. Therefore, we focus on the absolute value of $\delta S_d$. According to the maximum entropy principle (MEP), our results are the same for the time-dependent states (\ref{eqn:15}) and (\ref{eqn:16}).

Differential entropy of the density probability $\rho(\varphi,t)$ of the form of Eq. (\ref{eqn:13}) and normalized in the interval $\varphi\in[-\pi,\pi]$ takes the form of

\begin{align}\label{eqn:18}
\begin{split}
S_d(t)&= - \int_{-\pi}^{\pi} \rho(\varphi,t) \ln\rho(\varphi,t) d\varphi =\\&= - \ln\mathcal{N}(t) +
{\mathcal{N}(t)\over 2\sigma_{\varphi}^2(t)}\int_{-\pi}^{\pi}\varphi^2\exp(-\varphi^2/2\sigma_{\varphi}^2(t))d\varphi,\\
\mathcal{N}(t)&=\int_{-\pi}^{\pi} e^{-\varphi^2/2\sigma_{\varphi}^2(t)}d\varphi,
\end{split}
\end{align}

\noindent which is general for our considerations: placing $t=0$ follows the initial entropy form $S_d(0)$, while the standard deviation $\sigma_{\varphi}$ takes different forms for different CL processes and the classical counterpart \cite{JJD1}. On the use of Eq. (10), and Eq. (C.3) of Ref. \cite{JJD1}, where appear dimensionless $\sigma$s, the time-dependent standard deviation $\sigma_{\varphi}^2(t)$ takes the form of

\begin{align}\label{eqn:19}
\begin{split}
\sigma_{\varphi stand}^2(t) &= e^{-2n\gamma_{\circ}t} \left(\sigma_{\varphi}^2(0)\cos^2\omega t + {\sigma(0)\over 2n}\sin2\omega t + {\sigma_L^2(0)\over n^2}\sin^2\omega t\right)+\\
&+ {k_BT\over n I_{\circ}\omega^2}(1-e^{-2n\gamma_{\circ}t}),\\
\sigma_{\varphi markov}^2(t)& = e^{-2n\gamma_{\circ}t} \left(\sigma_{\varphi}^2(0)\cos^2\omega t + {\sigma(0)\over 2n}\sin2\omega t + {\sigma_L^2(0)\over n^2}\sin^2\omega t\right)+\\
&+ \left({k_BT\over n I_{\circ}\omega^2} - {\hbar^2\over 16 n I_{\circ}k_BT}\right)(1-e^{-2n\gamma_{\circ}t}),\\
\sigma_{\varphi decoh}^2(t) & = \sigma_{\varphi}^2(0)\cos^2\omega t + {\sigma(0)\over 2n}\sin2\omega t + {\sigma_L^2(0)\over n^2}\sin^2\omega t
+ {2\gamma_{\circ}k_B T\over n I_{\circ}\omega^2}t,
\end{split}
\end{align}

\noindent for the non-zero initial $\sigma_{\varphi}^2(0)$. For $t=0$ follows the initial standard deviation(s) and hence entropy $S_d(0)$.

\subsubsection{\label{sec:SectionIV31}The standard CL process}

The relative change of the differential entropy is found to increase with time.
For the underdamped regime we find a rich physical behavior. For the underdamped regime we choose $k_BT/I_{\circ}\omega^2=0.1$, and for the non-underdamped regime $k_BT/I_{\circ}\omega^2=10$.

For the initial values $\sigma_{\varphi}=\sqrt{0.1}, \sigma_L=\sqrt{10}$ and $\sigma_{\varphi}=\sqrt{0.1}, \sigma_L=\sqrt{5}$ and the underdamped regime, a decrease of $\vert\delta S\vert$ with the number of the propellers is found with the very small minimum values of 0.0005 and 0.0003, respectively, for $n=10$. Much larger values are found for other choices of the initial values yet with a {\it non-monotonic} behavior. For example, for the initial values $\sigma_{\varphi}=1, \sigma_L=\sqrt{5}$ and $\sigma_{\varphi}=\sqrt{0.9}, \sigma_L=\sqrt{1.1}$, the minimums 0.005 and 0.007 are found for $n=3$ and $n=1$, respectively.

For the non-underdamped regime, $\sigma_{\varphi}=\sqrt{0.1}, \sigma_L=\sqrt{5}$ and $\sigma_{\varphi}=\sqrt{0.1}, \sigma_L=\sqrt{10}$, a {\it non-monotonic} behavior is found: the minimums are found approximately 0.00002 for both cases, $n=7$ and $n=9$, respectively. Similar behavior is found for the initial values $\sigma_{\varphi}=1, \sigma_L=\sqrt{5}$ and $\sigma_{\varphi}=\sqrt{0.9}, \sigma_L=\sqrt{1.1}$ with the minimum approximately 0.003 for both choices of the initial conditions, for $n=2$ and $n=1$, respectively.

\begin{figure*}[!ht]
\centering
\includegraphics[width=0.4\textwidth]{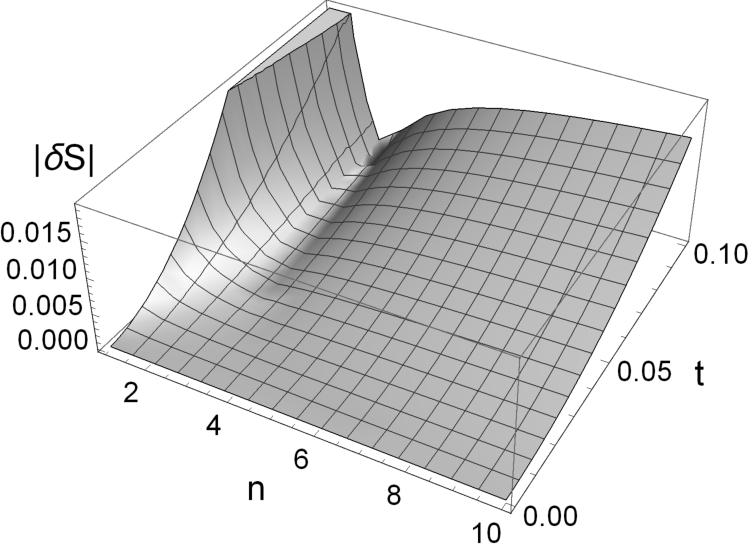}
\includegraphics[width=0.4\textwidth]{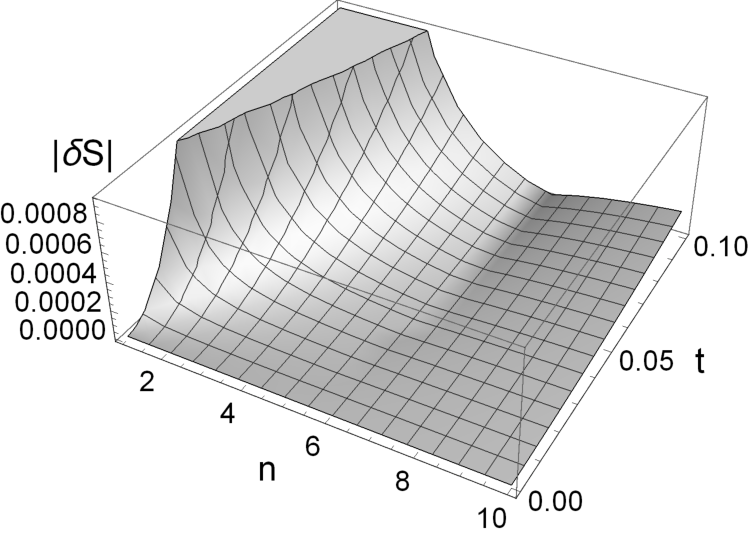}
\caption{Relative change of differential entropy for the standard CL processes (Left) Underdamped regime, initial conditions $\sigma_{\varphi}=1,\sigma_L=\sqrt{5}$, and (Right) Non-underdamped regime, with the initial conditions $\sigma_{\varphi}=\sqrt{0.1}, \sigma_L=\sqrt{5}$.}\label{Figure1}
\end{figure*}

In Figure~\ref{Figure1} we present the results for both under- and non-under-damped regimes, with different initial conditions to present existence of the local minimums, which is for $n=3$ for underdamped regime, while for the non-underdamped regime, the local minimum is found for $n=7$. Lower entropy change is found for the non-underdamped regime for certain initial values.

\subsubsection{\label{sec:SectionIV32}The Markovian-limit CL process}

The process is defined by very large temperature that corresponds to the underdamped regime. For all considered cases, the relative change of the differential entropy increases with time.

A simple behavior is found (for very high temperature we choose $k_BT/I_{\circ}\omega^2=1000$): the relative change of the differential entropy decreases with the increase of the number of propellers and the minimum values are found for $n=10$, but with much larger values than for the standard CL processes. For example, the smallest value of approximately 0.307 is found for the initial values $\sigma_{\varphi}=\sqrt{0.1}, \sigma_L=\sqrt{5}$. For this regime, the relative change of the differential entropy weakly depends on both the number of propellers as well as on the choice of the initial values.

\begin{figure*}[!ht]
\centering
\includegraphics[width=0.5\textwidth]{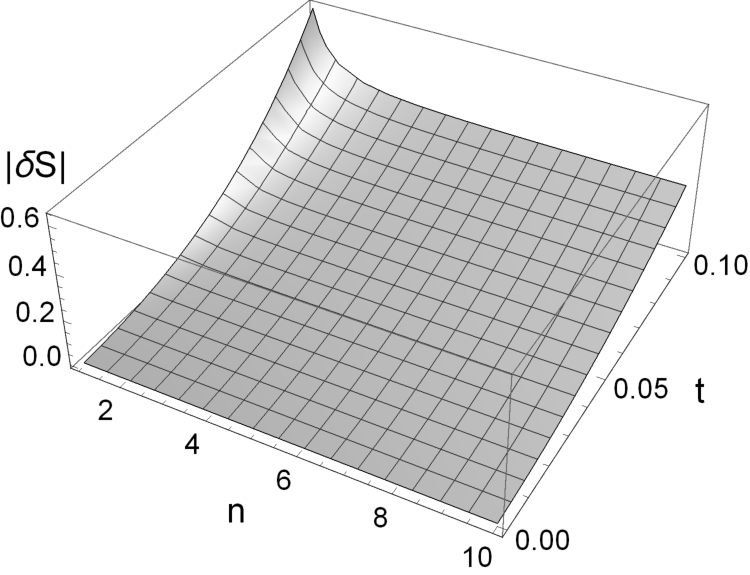}
\caption{Relative change of differential entropy for the Markovian process, with the initial conditions: $\sigma_{\varphi}=\sqrt{0.1},\sigma_L=\sqrt{10}$.}\label{Figure2}
\end{figure*}

\noindent Figure~\ref{Figure2} presents the case of the largest relative change of differential entropy for the considered initial conditions.

\subsubsection{\label{sec:SectionIV33}The decoherence-limit CL process}

The process is defined by very large moment of inertia $I_{\circ}$ for arbitrary values of other system/bath parameters. The relative change of the differential entropy increases in time for all considered cases.

Qualitatively the same behavior is found for both regimes, underdamped and non-underdamped, where $k_BT/I_{\circ}\omega^2=0.01$. The minimum values are found for $n=1$ for all the choices of the initial values, but with significant difference in the order of magnitude. For example, for the choice of the initial values $\sigma_{\varphi}=\sqrt{0.1}, \sigma_L=\sqrt{10}$, the relative change of the differential entropy is approximately 0.00024, also for $\sigma_{\varphi}=\sqrt{0.1}, \sigma_L=\sqrt{5}$.

\begin{figure*}[!ht]
\centering
\includegraphics[width=0.5\textwidth]{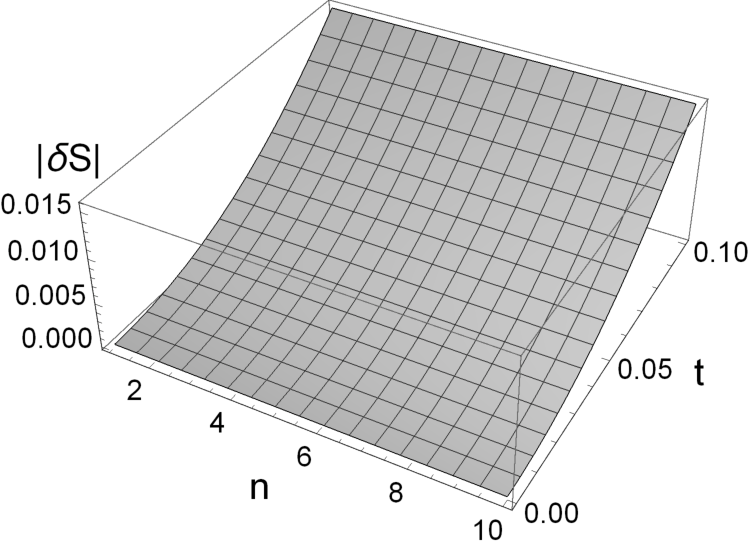}
\caption{Relative change of the differential entropy for the decoherence-limit process, with the initial values: $\sigma_{\varphi}=1,\sigma_L=\sqrt{5}$.}\label{Figure3}
\end{figure*}

Figure~\ref{Figure3} presents a case of relatively large relative change of differential entropy with the weak dependence on the number of propellers, thus exhibiting sensitivity to the choice of the initial values.

\subsection{\label{sec:SectionIV4}Linear entropy}

Linear entropy $S_L(t)=1-tr\hat\rho^2(t)$ is an approximation of the von Neumann entropy for the states close to a pure state. By definition, every entropy equals zero for a pure quantum state. Therefore, we are interested in the linear entropy for the mixed state Eq. (\ref{eqn:16})

\begin{equation}\label{eqn:20}
 \begin{split}
S_L(t)&=1-\int_{-\pi}^{\pi}\int_{-\pi}^{\pi} \rho^2(\varphi,\varphi',t) d\varphi d\varphi' =\\&
=1-\mathcal{N}^2\int_{-\pi}^{\pi}\int_{-\pi}^{\pi} \exp\left(-{1\over \sigma_{\varphi}^2(t)}\left({\varphi+\varphi'\over 2}\right)^2\right.-\\&-\left.{(I_{\circ}\omega\sigma_L(t))^2\over \hbar^2}(\varphi-\varphi')^2 \right) d\varphi d\varphi'.
\end{split}
\end{equation}

\noindent In Eq. (\ref{eqn:21}), $\sigma_{\varphi}^2(t)$ is presented by Eq. (\ref{eqn:19}), while $\sigma_L^2(t)$, partially borrowing from Eq. (\ref{eqn:10}), is of the following forms (neglecting $\gamma_{\circ}/\omega\ll 1$) \cite{JJD1}:

\begin{align} \label{eqn:21}
\begin{split}
\sigma_{L stand}^2(t)& = e^{-2n\gamma_{\circ}t} \left(\sigma_L^2(0)\cos^2\omega t - {n\over 2}\sigma(0)\sin2\omega t \right. +\\&  + \left. n^2\sigma_{\varphi}^2(0)\sin^2\omega t\right)
+ {n k_BT\over I_{\circ}\omega^2}(1-e^{-2n\gamma_{\circ}t}),\\
\sigma_{L markov}^2(t)& = e^{-2n\gamma_{\circ}t} \left(\sigma_L^2(0)\cos^2\omega t - {n\over 2}\sigma(0)\sin2\omega t + n^2\sigma_{\varphi}^2(0)\sin^2\omega t\right)
+\\&+ \left( {n k_BT\over I_{\circ}\omega^2}-{n\hbar^2\over 16I_{\circ}k_BT}\right)(1-e^{-2n\gamma_{\circ}t}),\\
\sigma_{L decoh}^2 & = {\sigma_L^2(0)}\cos^2 \omega t + n^2 \sigma_{\varphi}^2(0)\sin^2 \omega t - {n \sigma(0)\over 2}\sin 2\omega t + {2nk_BT\gamma_{\circ}\over I_{\circ}\omega^2}t.
\end{split}
\end{align}

Results presented below are numerically obtained with an emphasis on the cases found to have the relatively small entropy change. The values for $k_BT/I_{\circ}\omega^2$ and $I_{\circ}$ are the same as for the differential entropy, for every regime separately.

\subsubsection{\label{sec:SectionIV41}The standard CL process}

For the underdamped regime, the relative change of the linear entropy increases with time, but for the non-underdamped regime, it is found that the entropy change can decrease, depending on the number of  blades and the initial conditions.

For the underdamped regime, the relative change of the linear entropy is found to increase with time, but with non-monotonic behavior for some choices of the initial conditions. For example, for the initial values $\sigma_{\varphi}=1, \sigma_L=\sqrt{5}$ and $\sigma_{\varphi}=\sqrt{0.9}, \sigma_L=\sqrt{1.1}$, the minimum values 0.0005 and 0.002 are found for $n=2$ and $n=1$, respectively. Much larger values are found for the initial values $\sigma_{\varphi}=\sqrt{0.1}, \sigma_L=\sqrt{10}$ and $\sigma_{\varphi}=\sqrt{0.1}, \sigma_L=\sqrt{5}$ with the nonmonotonic behavior: the values are 0.016 and 0.039 for $n=8$ and $n=5$, respectively. Similar behavior is found for the non-underdamped regime.

\begin{figure*}[!ht]
\centering
\includegraphics[width=0.4\textwidth]{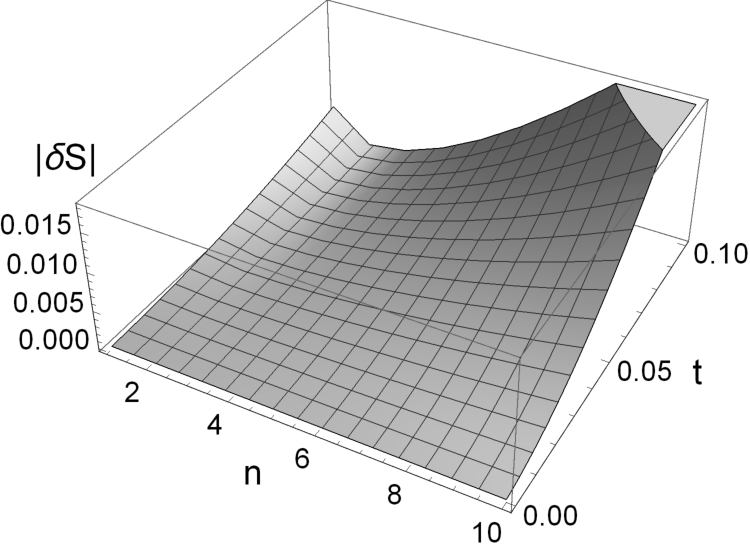}
\includegraphics[width=0.4\textwidth]{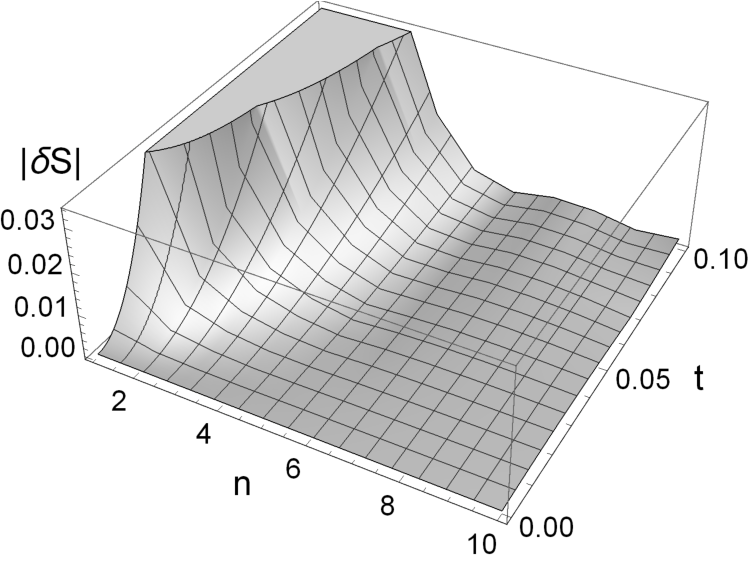}
\caption{Relative change of linear entropy for the standard CL processes (Left) Underdamped regime, initial conditions $\sigma_{\varphi}=1,\sigma_L=\sqrt{5}$, and (Right) Non-underdamped regime, with the initial conditions $\sigma_{\varphi}=\sqrt{0.1}, \sigma_L=\sqrt{5}$.}\label{Figure4}
\end{figure*}

From Figure~\ref{Figure4} it follows existence of the local minimums $n=2$ for the underdamped and two local minimums for $n=6$ and $n=9$ for the non-underdamped regime.

\subsubsection{\label{sec:SectionIV42}The Markovian-limit CL process}

The relative change of the linear entropy increases with time for all the cases considered.

Like for the differential entropy, we find the relatively large values, e.g. 0.17 for the initial values $\sigma_{\varphi}=\sqrt{0.9}, \sigma_L=\sqrt{1.1}$ for $n=1$. Interestingly, non-monotonic behavior is found for the initial values $\sigma_{\varphi}=\sqrt{0.1}, \sigma_L=\sqrt{10}$, where the minimum of 0.463 is found for $n=2$.

\begin{figure*}[!ht]
\centering
\includegraphics[width=0.5\textwidth]{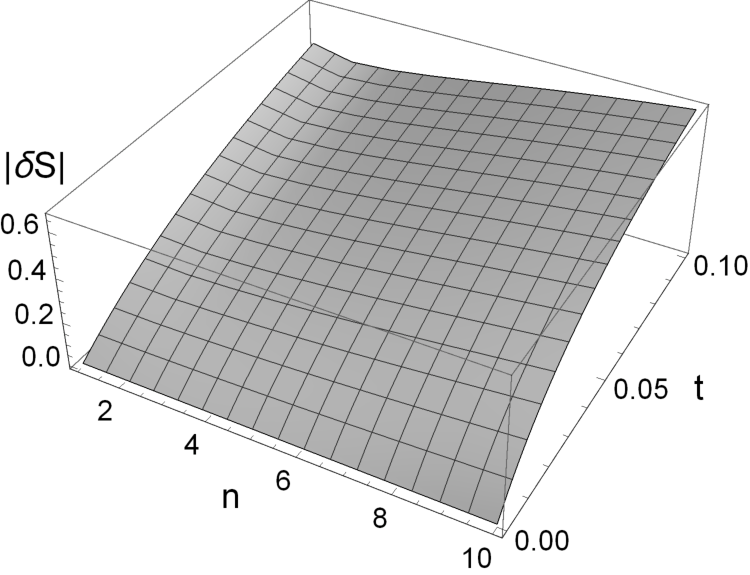}
\caption{Relative change of linear entropy for the Markovian process, with the initial conditions: $\sigma_{\varphi}=\sqrt{0.1},\sigma_L=\sqrt{10}$.}\label{Figure5}
\end{figure*}

Figure~\ref{Figure5} presents a specific case for the Markovian regime, where appears existence of a local minimum for $n=2$; for larger number of propellers, there is a sudden increase of the relative change of entropy in time.

\subsubsection{\label{sec:SectionIV43}The decoherence-limit CL process}

The process is defined by a very large moment of inertia $I_{\circ}$ for arbitrary values of other system/bath parameters. For all the considered cases, the relative change of linear entropy increases with time.

For both underdamped and non-underdamped regimes, qualitatively the same behavior is found. Typically, large values of the relative change of the linear entropy appear, except for the initial values $\sigma_{\varphi}=1, \sigma_L=\sqrt{5}$, for which the change of approximately 0.283 is found for $n=1$. Like for the differential entropy, it is found weak dependence on the number of blades and the regime.

\begin{figure*}[!ht]
\centering
\includegraphics[width=0.5\textwidth]{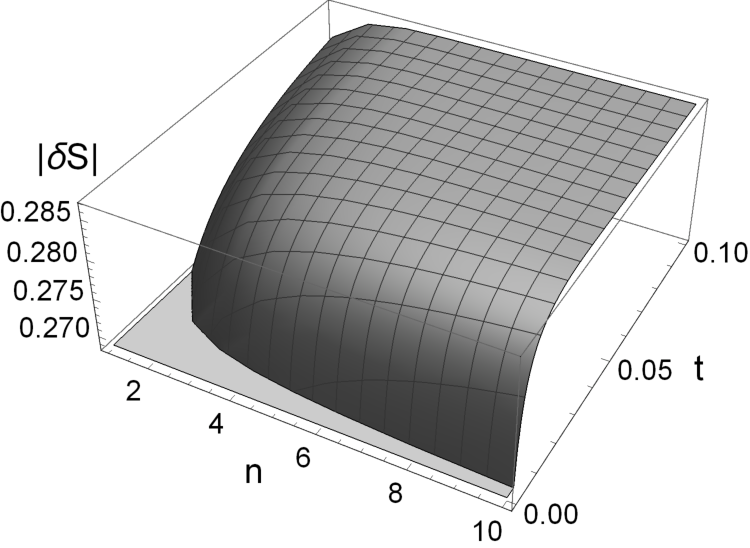}
\caption{Relative change of the linear entropy for the decoherence-limit process, with the initial values: $\sigma_{\varphi}=1,\sigma_L=\sqrt{5}$.}\label{Figure6}
\end{figure*}

In Figure~\ref{Figure6} we present a choice of the initial conditions, for which relative change linear entropy is not very large, approximately in between 0.283 and 0.287.

\subsection{\label{sec:SectionIV44}Overview}

Differential and linear entropy provide different insights into the system. If one is interested in controlling the angle of rotation, they should use differential entropy. If control of the rotator state is of interest, then linear entropy should be used. Both entropies help distinguish between various CL processes. However, this isn't true for the information-backflow criterion of (non)Markovianity \cite{BROJER}.

In the decoherence-limit CL process, both entropies show weak dependence on the regime and the number of propellers, but they are highly influenced by initial conditions. The Markovian process results in the largest relative change for both entropies. Significant changes are noted in linear entropy during the decoherence process. Typically, the change in linear entropy is greater than that of differential entropy.

While the time behavior is generally a monotonic increase for every $n$, we observe {\it non-monotonic} dependence on the number of blades, $n$, in certain cases. This is especially true for standard CL processes (under- and non-underdamped)  and for linear entropy regarding the Markovian process. For instance, in Figure~\ref{Figure4}(Right), a small "hill" separates local minimums
near $n = 5$ and $n = 9$, highlighting the local minimums of dynamical stability.
This suggests that there are no simple, general rules or recipes for achieving relatively stable rotations.

\section{\label{sec:SectionV}Discussion}

In line with the MEP, we create a mixed and a pure state as candidate states for achieving proper dynamical stability in molecular cogwheels exposed to an external harmonic field. The results in Section~\ref{sec:SectionIV} show sensitivity to variations in CL processes and parameter choices, but do not provide simple rules for achieving relatively stable dynamics in molecular rotators.

We consider two entropies in this paper, both regarding the angle of rotation as the rotator observable. In some cases, the linear entropy yields values that are orders of magnitude larger (see, for example, Figure~\ref{Figure1}(Right) compared to Figure~\ref{Figure4}(Right), or Figure~\ref{Figure3} compared to Figure~\ref{Figure6}). This makes the linear entropy a more stringent criterion for rotational stability. However, if the laboratory procedure involves manipulating the {\it angle} of rotation--such as by applying an external electric field to polar molecules--the results for the differential entropy should be used. On the other hand, if the procedure involves manipulating the rotator {\it state}, the results for the linear entropy are of interest.

Many scenarios exist for every process and regime, corresponding to a chosen value of the ratio $F=k_BT/I_{\circ}\omega^2$ (Section~\ref{sec:SectionIV}). In other words, numerous combinations of ($T, I_{\circ},\omega$) values correspond to a given $F$. The frequency $\omega$ then determines the physical units and the time scale $t\ll\omega$.

Section~\ref{sec:SectionIV} shows a {\it non-monotonic} dependence of entropy change on the number of blades, $n$. This is illustrated in Figure~\ref{Figure1}, Figure~\ref{Figure4}(Right), and Figure~\ref{Figure5}. Notably, Figure~\ref{Figure1}(Right) reveals a small "hill" between the two local minimums at $n = 5$ and $n = 9$. These minimums indicate the "best" choices for stable rotation. For instance, Figure~\ref{Figure4}(Right) suggests using the standard CL process in a non-underdamped regime with initial values of $\sigma_{\varphi}=\sqrt{0.1}$ and $\sigma_L=\sqrt{0.5}$ for $n = 5$ blades. This pattern holds true for the other figures as well.

The entropy change exhibits rich behavior, differing from the results in Ref. \cite{JJD1}, which reports monotonic behavior for the standard deviation of the angle in a harmonic rotator. In Ref. \cite{JJD1}, the best stability, shown by the lowest $\sigma_{\varphi}$, occurs at $n = 10$. As noted throughout this paper, deciding which stability criterion--standard deviation or entropy dynamics--is more useful must be experimentally decided, regardless of whether the molecular rotator starts in a mixed or pure state.

In regard to the initial pure states, we discover some interesting possibilities for achieving relatively stable rotation. In realistic physical situations, the equalities in Eq. (\ref{eqn:A5}) are only approximately satisfied. These expressions may only apply to the standard CL process in Eq. (\ref{eqn:novo1}). The condition $2k_BT/\hbar\omega\gg 1$ for the Markovian case contrasts with the approximate pure-state condition $2k_BT/\hbar\omega= 1$ (cf. Appendix \ref{AppendixA}), while, the pure state condition $nI_{\circ}=\hbar^2/4k_BT$ leads to the approximate $n\gamma_{\circ}/2$ instead of the large constant $2n^2I_{\circ}\gamma_{\circ}k_BT/\hbar^2$ appearing in the CL decoherence limit $I_{\circ}\gg 1$. Therefore, the purity loss for the Markovian and decoherence-limit CL processes cannot be based on the purity conditions of Appendix~\ref{AppendixA}. Different scenarios are possible for implementing these findings. For example, we can choose the rotator parameters $nI_{\circ}$ and $\sigma_{\varphi}^2$ to approximately satisfy Eq. (\ref{eqn:A5}) for the choices of the external field frequency $\omega\approx\hbar/2nI_{\circ}\sigma_{\varphi}\sqrt{2}$ and the temperature $k_BT\approx\hbar^2/4nI_{\circ}\sigma_{\varphi}^2$. There are other similar choices for the state preparation (the value of $\sigma_{\varphi}$) and the physical scenario (the values of $T$ and $\omega$) with the freedom to choose the number $n$ and the average inertia $I_{\circ}$ of the propeller blades to approximately fulfill the conditions of relatively stable molecular rotations - with the satisfied relations $k_BT/nI_{\circ}\approx\omega^2$ and $L\equiv\langle\hat L_z\rangle=0$.

Let's conduct quantitative tests on the scenarios. All scenarios should follow realistic assumptions (in SI units \cite{MomentOfInertiaValues}) of $k_BT\sim 10^{-21}$J and $I_{\circ}\sim 10^{-46}-10^{-42}$kgm$^2$. First, we find $\omega\sim 10^{12}-10^{13}$s$^{-1}$, $nI_{\circ}\sim 10^{-46}$kgm$^2$, and $\sigma_{\varphi}=1$. Alternatively, we can use $\sigma_{\varphi}^2=0.1$, as discussed in Section~\ref{sec:SectionIV}. This gives us values that satisfy Eq. (\ref{eqn:A5}) with $nI_{\circ}\sim 10^{-44}$kgm$^2$, $\omega\sim 10^{12}$s$^{-1}$ and $k_BT\sim 10^{-21}$J. Thus, a range of $\sigma_{\varphi}$ values supports our analysis in Section~\ref{sec:SectionIV2} and some experimental evidence \cite{MomentOfInertiaValues,Ribeto}.

To back these estimates, we reference the field of intense laser alignment of molecules \cite{Molecularwavepackets1,Molecularwavepackets2}. Molecular alignment techniques can create broad wave-packet rotational states, even for polyatomic molecules \cite{Molecularwavepackets1}, starting from a thermal state of the gas \cite{Molecularwavepackets2}. For the body-fixed azimuthal angle, panel (c) in Figure 5 of Ref. \cite{Molecularwavepackets1} shows a high probability for small angle values, aligning with our constraint $\sigma_{\varphi}(t)\le 1$. The chosen external laser fields are approximately resonant with electronic or rovibrational transition frequencies, roughly in the range of $10^{10}-10^{16}$s$^{-1}$, consistent with our estimates. The temperature range is $k_BT\sim 10^{-22}-10^{-21}$J, again aligning with our findings. The low rotational temperatures relate to the task of aligning the molecules, which is not the focus of this paper.

\section{\label{sec:SectionVI}Conclusion}

We propose and construct an initial state, either pure or mixed, for a harmonic molecular cogwheel, which yields a relatively low entropy change for both differential and linear entropy in the main Caldeira-Leggett processes. By making realistic choices of system and initial-state parameters, we demonstrate that this goal can be achieved. Different combinations of these values represent different physical scenarios for attaining a relatively low entropy change in all constructed initial states, processes, and dynamical regimes of the molecular cogwheel. The practical usefulness of our findings can only be determined experimentally.

\bigskip

\noindent {\bf Acknowledgements} This research is funded by the Ministry of Education and Ministry of Science, Technological Development and Innovation, Republic of Serbia, Grants no. 451-03-137/2025-03/200124 and 451-03-137/2025-03/200122.

\bigskip

\appendix

\section{\label{AppendixA}Approximate stationary states}

In the ''position'' representation, the standard CL master equation (\ref{eqn:novo1}) is of the well known form \cite{Brojer} that is here given for the harmonic rotator:

\begin{equation}\label{eqn:A1}
\begin{split}
 & {\partial\rho(\varphi,\varphi',t)\over \partial t}= \left[
{\imath\hbar\over 2n I_{\circ}}\left({\partial^2\over\partial \varphi^2}-{\partial^2\over\partial \varphi^{'2}}\right)
-{\imath n I_{\circ}\omega^2\over 2\hbar}(\varphi^2-\varphi^{'2})
\right.-\\&- \left.\gamma(\varphi-\varphi')\left({\partial\over\partial\varphi}-
{\partial\over\partial\varphi'} \right)
\right]\rho(\varphi,\varphi',t)
-{2nI_{\circ}\gamma k_BT\over\hbar^2}(\varphi-\varphi')^2\rho(\varphi,\varphi',t).
\end{split}
\end{equation}

The time-independent kernel of the state Eq. (\ref{eqn:16}) gives an exact stationary state for the process Eq. (\ref{eqn:novo1}).
Providing the very high temperature ($T\gg 1$), it can be easily shown that the approximate relations apply for the Markovian-limit process where appears an additional term:

\begin{equation}\label{eqn:A2}
\begin{split}
{\partial\rho(\varphi,\varphi',t)\over \partial t}& = \left[
{\imath\hbar\over 2n I_{\circ}}\left({\partial^2\over\partial \varphi^2}-{\partial^2\over\partial \varphi^{'2}}\right)
-{\imath n I_{\circ}\omega^2\over 2\hbar}(\varphi^2-\varphi^{'2})
\right.-\\&- \left.\gamma(\varphi-\varphi')\left({\partial\over\partial\varphi}-{\partial\over\partial\varphi'} \right) \right]\rho(\varphi,\varphi',t)-\\&
-\left[{2nI_{\circ}\gamma k_BT\over\hbar^2}(\varphi-\varphi')^2-{\gamma\hbar^2\over 8n I_{\circ}k_BT} \left({\partial\over\partial\varphi} - {\partial\over\partial\varphi'} \right)^2\right]\rho(\varphi,\varphi',t),
\end{split}
\end{equation}

\noindent but {\it not} for the decoherence-limit process of the following form:

\begin{equation}\label{eqnA3}
\begin{split}
{\partial\rho(\varphi,\varphi',t)\over \partial t}& = \left[
{\imath\hbar\over 2n I_{\circ}}\left({\partial^2\over\partial \varphi^2}-{\partial^2\over\partial \varphi^{'2}}\right)
-{\imath n I_{\circ}\omega^2\over 2\hbar}(\varphi^2-\varphi^{'2})
\right]\rho(\varphi,\varphi',t)-\\&-{2nI_{\circ}\gamma k_BT\over\hbar^2}(\varphi-\varphi')^2\rho(\varphi,\varphi',t).
\end{split}
\end{equation}

Interestingly, the (time-independent) pure state Eq.(\ref{eqn:15}) can be recognized as an exact stationary state for the CL process Eq. (\ref{eqn:novo1}). For completeness, we give the main steps in the proof of this claim. The first and the second derivatives of the state Eq. (\ref{eqn:15}) are readily obtained:

\begin{align}\label{eqn:A4}
\begin{split}
{\partial\rho(\varphi,\varphi',t)\over \partial \varphi}&= \left(-{\varphi\over 2\sigma_{\varphi}^2}+{\imath\over\hbar}L\right)\rho(\varphi,\varphi',t),\\
{\partial\rho(\varphi,\varphi',t)\over \partial \varphi'} &= \left(-{\varphi'\over 2\sigma_{\varphi}^2}-{\imath\hbar}L\right)\rho(\varphi,\varphi',t),\\
{\partial^2\rho(\varphi,\varphi',t)\over \partial \varphi^2} &= \left(-{1\over 2\sigma_{\varphi}^2}+\left({\varphi\over 2\sigma_{\varphi}^2}-{\imath\over\hbar}L\right)^2
\right)\rho(\varphi,\varphi',t),\\
{\partial^2\rho(\varphi,\varphi',t)\over \partial \varphi^{'2}}& = \left(-{1\over 2\sigma_{\varphi}^2}+\left({\varphi'\over2\sigma_{\varphi}^2}+{\imath\over\hbar}L\right)^2
\right)\rho(\varphi,\varphi',t).
\end{split}
\end{align}

\noindent Substituting those terms in Eq. (\ref{eqn:A1}) after some simple algebra follow the conditions for the pure state dynamics $\partial\rho(\varphi,\varphi',t)/\partial t=0$:
\begin{equation*}\label{eqn:A5}
{\hbar\over 2n I_{\circ}}{1\over 4\sigma_{\varphi}^2}={n I_{\circ}\omega^2\over 2\hbar},
\quad{\gamma\over\hbar}L=0,\qquad {L\over 2n I_{\circ} \sigma_{\varphi}^2}=0,\quad
{1\over 2\sigma_{\varphi}^2}={2n I_{\circ}k_BT\over\hbar^2}.\tag{A.5}
\end{equation*}

\noindent Bearing in mind that (for the process Eq. (\ref{eqn:novo1})) $\sigma_{\varphi}^2=k_BT/n I_{\circ}\omega^2$, for a given choice of the temperature $T$ and the mean of the angular momentum $L\equiv \langle\hat L_z\rangle=0$, the equalities $n I_{\circ}=\hbar^2/4k_BT$ and $\omega=2k_BT/\hbar$ give rise to $\sigma_{\varphi}^2=1$ and to the {\it exact} equalities in Eq. (\ref{eqn:A5}). Nevertheless, a word of caution is in order: arbitrary low temperature would require physically unreasonable large moment of inertia $I_{\circ}$ and very small frequency $\omega$.

Now it is rather obvious that, for very high temperature, due to the last term in Eq. (\ref{eqn:A2}), the pure state (\ref{eqn:15}) is approximately stationary for the Markovian, but not for the decoherence-limit of the CL process.

\section{\label{AppendixB}Characteristics of the ansatz probability distribution}

\begin{figure}[h!]
\centering
\includegraphics[width=0.5\textwidth]{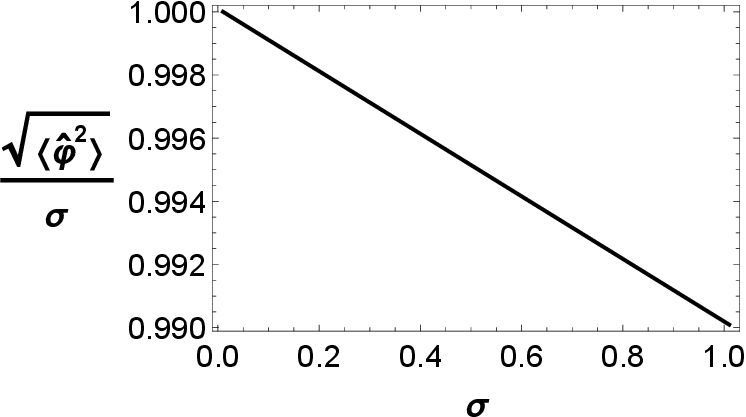}
\caption{The ratio $\sqrt{\langle\hat\varphi^2\rangle}/\sigma$. The match is excellent for $\sigma < 1.2$.}\label{FigEqualStandardDeviations}
\end{figure}

First, we show that the standard deviation $\sigma_{\varphi}$ appearing in Eqs. (\ref{eqn:13})-(\ref{eqn:16}) can remain approximately valid also for the integration over the interval $\varphi\in[-\pi,\pi]$. That is, we show that (for the angle zero mean) $\langle\hat\varphi^2\rangle=\mathcal{N}\int_{-\pi}^{\pi}\varphi^2 \exp(-{\varphi^2\over 2\sigma_{\varphi}^2}) d\varphi \approx \sigma_{\varphi}^2=\int_{-\infty}^{\infty}\varphi^2\rho(\varphi)d\varphi$, where $\rho(\varphi)$ is given by Eq. (\ref{eqn:13}). Figure~\ref{FigEqualStandardDeviations} exhibits validity of $\langle\varphi^2\rangle\approx\sigma_{\varphi}^2$ for all the instantaneous values of $\sigma_{\varphi}(t)\le 1$. Let us emphasize that
the normalization factor $\mathcal{N}$ is given by Eq. (\ref{eqn:B1}). Interestingly, the range of $\sigma_{\varphi}\in [0.01,1]$ is in accord with the previously obtained values for the angle standard deviation \cite{JJD1,JJD2} even for the long time intervals.

\begin{figure}[h!]
\centering
\includegraphics[width=0.6\textwidth]{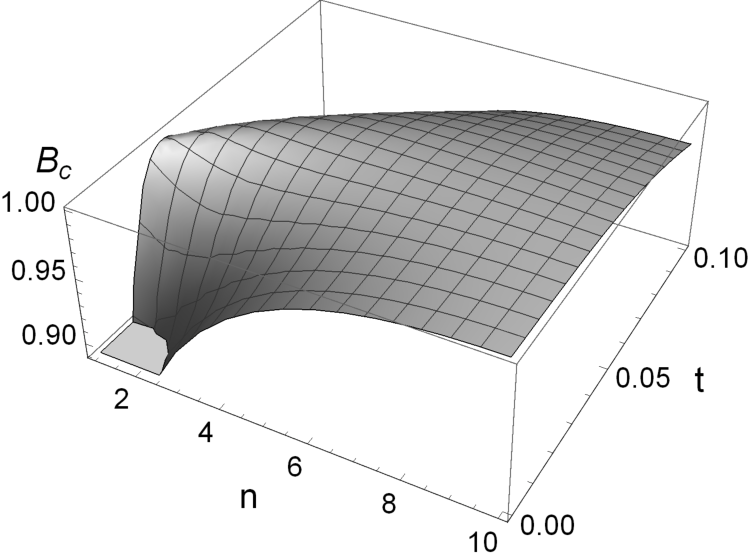}
\caption{Overlap of the time-dependent and time-independent rotator states. The smallest value of approximately $0.3$ is for $n=1$.}\label{FigStateOverlap}
\end{figure}

Next, we investigate the distance between the time-independent $\rho(\varphi)=\mathcal{N}\exp(-\varphi^2/2\sigma_{\varphi}^2)$ and the time-dependent (ansatz) probability distribution $\rho(\varphi,t)=\mathcal{M}\exp(-\varphi^2/2\sigma_{\varphi}^2(t))$, where $\sigma_{\varphi}^2(t)$ is for the process Eq. (\ref{eqn:novo1}). We choose the so-called Bhattacharyya distance measure (Bhattacharyya coefficient), $B_c=\int dx \sqrt{f(x)g(x)}$, for the two probability distributions $f(x)$ and $g(x)$; $0\le B_c\le 1$, where $B_c=1$ is for the zero distance of $f(x)=g(x)$. The normalization coefficients and the integral are readily calculated,

\begin{align}\label{eqn:B1}
\begin{split}
\mathcal{N}&=\int_{-\pi}^{\pi}\exp\left(-{\varphi^2\over 2\sigma_{\varphi}^2}\right)d\varphi = \left(\sigma_{\varphi} \sqrt{2\pi} {\rm Erf}[\pi/\sigma_{\varphi}\sqrt{2}]\right)^{-1},\\
\mathcal{M}&=\int_{-\pi}^{\pi}\exp\left(-{\varphi^2\over 2\sigma_{\varphi}^2(t)}\right)d\varphi =\\&=
\left(\sigma_{\varphi}(t) \sqrt{2\pi} {\rm Erf}[\pi/\sigma_{\varphi}(t)\sqrt{2}]\right)^{-1},\\
I&=\int_{-\pi}^{\pi}\exp(-\alpha\varphi^2) d\varphi = {\sqrt{\pi} {\rm Erf}[\pi\sqrt{\alpha}]\over \sqrt{\alpha}},\\
\alpha & = {1\over\sigma_{\varphi}^2}+{1\over\sigma_{\varphi}^2(t)},
\end{split}
\end{align}

\noindent with the Bhattacharyya coefficient $B_c=\mathcal{N}\mathcal{M}I$ presented in Figure~\ref{FigStateOverlap}; by ''Erf'' we assume the well-known Glaisher ''error function''.

\noindent From Figure~\ref{FigStateOverlap} we can see that the two probability distributions become mutually comparable already for a rather short time interval, especially for the larger propellers and larger initial values of the standard deviations $\sigma_{\varphi}(0)$ and $\sigma_L(0)$.  The initial numerical discrepancy is due to the very small, arbitrarily chosen initial value of $\sigma_{\varphi}(t=0)=0.1$. Similar results are obtained for the Markovian-, but (expectably \cite{JJD1,JJD2}) not for the decoherence-limit of the CL process.

\end{document}